\documentclass[fleqn,12pt,twoside]{article}
\usepackage{espcrc1}

\usepackage{epsfig}
\usepackage[figuresright]{rotating}

\def\gtrless{\raise2.5pt\hbox{$>$}\llap{\lower2.5pt\hbox{$<$}}}

\def\gtrapprox{\,\raise2.5pt\hbox{$>$}\llap{\lower2.5pt\hbox{$\sim$}}\,}

\newcommand{\bsq}[1]{\begin{subequations}\label{#1}}

\newcommand{\esq}{\end{subequations}}

\newcommand{\beq}{\begin{equation}}

\newcommand{\eeq}{\end{equation}}

\newcommand{\beqa}[1]{\begin{eqnarray}\label{#1}}

\newcommand{\eeqa}{\end{eqnarray}}

\newcommand{\gd}{\dot{\gamma}}

\def\lessapprox{\raise2.5pt\hbox{$<$}\llap{\lower2.5pt\hbox{$\approx$}}}

\newcommand{\AmS}{{\protect\the\textfont2
  A\kern-.1667em\lower.5ex\hbox{M}\kern-.125emS}}

\title{Schematic Mode Coupling Theories for Shear Thinning, Shear 
Thickening, and  Jamming}  

\author{M. E. Cates and C. B. Holmes\\ 
School of Physics, The University of
Edinburgh,\\ JCMB Kings Buildings, Edinburgh EH9 
3JZ, Scotland\\
\vspace{0.5cm}
M. Fuchs and O. Henrich\\ 
Fachbereich Physik, Universit\"at Konstanz,\\
D-78457 Konstanz, Germany}  
\begin{document}
% typeset front matter
\maketitle
\begin{abstract}
Mode coupling theory (MCT) appears to explain several, though
not all, aspects of the glass transition in colloids 
(particularly when short-range attractions are present). Developments
of MCT, from rational foundations in statistical mechanics, account
qualitatively for nonlinear flow behaviour such as the yield stress of
a hard-sphere  
colloidal glass. Such theories so far only predict shear thinning
behaviour, whereas in real colloids both shear thinning and shear
thickening can be found. The latter observation can, however, be
rationalised by postulating an MCT vertex that is not only a
decreasing function of strain rate (as found from first principles)
but also an increasing function of stress. Within a highly simplified,
schematic MCT model this can lead not only to discontinuous shear
thickening but also to complete arrest of a fluid phase under the
influence of an external stress (`full jamming').  
\end{abstract}

\section{INTRODUCTION}

\subsection{Arrest in Colloidal Fluids}

Colloidal fluids can be studied relatively easily by light scattering
\cite{book,pine}. This allows one to measure the dynamic structure
factor $S(q,t_1-t_2) = \langle\rho({\bf q},t_1)\rho({\bf
  -q},t_2)\rangle/N$ and also the static one, $S(q)= S(q,0)$. Here
$\rho({\bf r},t) = \sum_i\delta({\bf r}_i(t)-{\bf r})- N/V$; this is
the real space particle density (with the mean value subtracted), and
$\rho({\bf q},t)$ is its Fourier transform. For particles of radius
$a$ with short-range repulsions, $S(q)$ exhibits a peak at a value
$q^*$ with $q^* a = {\cal O}(1)$. The dynamic structure factor
$S(q,t)$, at any $q$, decays monotonically from $S(q)$ as $t$
increases. In an ergodic colloidal fluid, $S(q,t)$ decays to zero
eventually: all particles can move, and the density fluctuations have
a finite correlation time. In an arrested state, which is nonergodic,
this is not true. Instead the limit $S(q,\infty)/S(q) = f(q)$ defines
the {\em nonergodicity parameter}. The presence of nonzero $f(q)$
signifies frozen-in density fluctuations. Although $f(q)$ is strongly
wavevector dependent, it is common to quote only $f(q^*)$ \cite{kob}.  
The above formulas assume time-translation 
invariance; nonergodic systems can violate this 
(showing e.g. aging phenomena) in which case
$S(q,t_1-t_2)$ as defined above must be written
$S(q,t_1,t_2)$ with two time arguments.

In many colloidal materials the effective interparticle interaction
$u(r)$ comprises a hard sphere repulsion, operative at separation
$2a$, perhaps combined with an attraction at larger distance. (For
simplicity one can imagine a square well potential of depth $\epsilon$
and range $\xi a$, with $\xi < 1$ typically.) Colloidal fluids of this
type are found to undergo nonergodicity transitions into two different
broad classes of arrested nonequilibrium states. One is the colloidal
glass, in which arrest is caused by the imprisonment of each particle
in a cage of neighbours. This occurs even for $\epsilon = 0$
(i.e. hard spheres) at volume fractions above about $\phi \equiv 4\pi
a^3N/3V \simeq 0.58$. The nonergodicity parameter for the  
colloidal glass obeys $f(q^*)\simeq 0.8$.  The second arrested state
is called the colloidal gel. Unlike the repulsive glass, the arrest
here is driven by attractive interactions, resulting in a bonded,
network-type structure. Such gels can be unambiguously found, for
short range attractions, whenever $\beta\epsilon \gtrapprox
5-10$. (Here $\beta \equiv 1/k_BT$.) Hence it is not necessary that
the local bonds are individually irreversible (this happens,
effectively, at $\beta\epsilon \gtrapprox 15-20$); and when they are
not, the arrest is a collective, not just a local, phenomenon. It is
found experimentally that for colloidal gels, $f(q^*) \gtrapprox 0.9$,  
which is distinctly different from the colloidal 
glass. The arrest line for gel formation slices across the equilibrium
phase diagram (e.g., plotted on the $(\phi,\beta\epsilon)$ plane),
and, depending on $\xi$, parts of it lie within two phase
regions. This, alongside any metastable gas-liquid phase boundary that
is present, can lead to a lot of interesting kinetics
\cite{poon,kroy}, in which various combinations of phase separation
and gelation lead to complex microstructures and time evolutions.

\subsection{Mode Coupling Theory (MCT)}

We do not review MCT in detail here. One widely used
form of the theory \cite{goetze} is based on projection
methods. However, in a stripped down version (see
e.g. \cite{ramaswamy,kk}) the resulting equations can be viewed as a
fairly standard one-loop selfconsistent approximation to a dynamical
theory for the particle density field.

We take $\beta = 1$, bare particle diffusivities $D_0 = 1$, and start
from the overdamped Langevin equations $\dot {\bf r}_i = {\bf F}_i +
{\bf f}_i$ for independent particles of unit diffusivity subjected to
external forces ${\bf F}_i$ and random forces ${\bf f}_i$. One
proceeds by a standard route to a Smoluchowski equation $\dot \Psi =
\Omega \Psi$ for the $N$-particle distribution function $\Psi$, with
evolution operator $\Omega = \sum_i\nabla_i.(\nabla_i-{\bf F}_i)$. Now
take the forces ${\bf F}_i$ to originate (via ${\bf F}_i = -\nabla_i
H$) from an interaction Hamiltonian  
\beq
H = -\frac{1}{2}\int d^3{\bf r}d^3{\bf r'} \rho({\bf r})\rho({\bf
  r'})c(|{\bf r}-{\bf r'}|) 
\label{hamiltonian}
\eeq
where $N c(q) = V[1-S(q)^{-1}]$. This is a harmonic expansion in
density fluctuations; $c(q)$ is the direct correlation function, and
this form ensures that $S(q)$ is recovered in equilibrium. We neglect
solvent mediated dynamic forces (hydrodynamic couplings). Also, in
principle these couplings mean that the noise in the Langevin equation
should be correlated between particles, in contrast to the independent
white noise assumed here. In addition we neglect anharmonic terms in
$H$; to regain the correct higher order density correlators (beyond
the two point correlator $S(q)$) in equilibrium, these terms would
have to be put back.

From the Smoluchowski equation (or the corresponding nonlinear
Langevin equation for the density $\rho({\bf
  r})$\cite{ramaswamy,paris}), one can derive a hierarchy of equations
of motion for correlators such as $S(q,t)$, more conveniently
expressed via $\Phi(q,t) \equiv S(q,t)/S(q)$. Factoring arbitrarily
the four-point correlators that arise in this hierarchy into products
of two $\Phi$'s, one obtains a closed equation of motion for the two
point correlator 
\beq
\dot\Phi(q,t) + \Gamma(q)\left[ \Phi(q,t) + \int_0^t
  m(q,t-t')\dot\Phi(q,t')dt'\right] = 0 
\label{correlator}
\eeq
where $\Gamma(q) = q^2/S(q)$ is an initial decay rate, and the memory
function obeys 
\beq
m({\bf q},t)= \sum_{{\bf k}} V_{{\bf q},{\bf k}}\Phi({\bf
  k},t)\Phi({\bf k-q},t) 
\label{memory}
\eeq
with the vertex 
\beq
V_{{\bf q},{\bf k}} = \frac{N}{2V^2q^4}S(q)S(k)S(|{\bf k-q}|)[{\bf
  q}.{\bf k}c(k) + {\bf q}.( 
{\bf k}-{\bf q}) c(|{\bf k-q}|)]^2
\label{vertex}
\eeq

Equations \ref{correlator}-\ref{vertex}, which are slightly simpler
than the ones used in molecular glasses because of the justified
neglect of inertial terms in an overdamped environment, completely
define the MCT as usually applied in colloidal systems \cite{goetze}. 

The MCT equations exhibit a bifurcation that corresponds to a sudden
arrest transition, upon smooth variation of either the density $\phi$
or any interaction parameters that control $c(q)$ (equivalently,
$S(q)$). Here the nonergodicity parameters $f(q)$, suddenly jump (for
all $q$ at once) from zero to nonzero values.  Near this (on the
ergodic side, which is always the direction MCT approaches from),
$\Phi(q,t)$ develops interesting behaviour. Viewed as a function of
time,  it decays onto a plateau of height $f(q)$, stays there for a
long time, and then finally decays again at very late times. The two
decays are called $\beta$ and $\alpha$ respectively. Upon crossing the
bifurcation, their relaxation times diverge smoothly with the
parameters; upon crossing the locus of this divergence, $f(q) \equiv
S(q,\infty)$ jumps discontinuously from zero to a finite value.

\section{MCT \& DYNAMIC HETEROGENEITY}
\label{assisted}

It is interesting to compare the MCT approach with
the concept of dynamical heterogeneity 
and/or `assisted dynamics' (e.g. \cite{garrahan}). 
MCT ignores locally activated processes 
but treats collective density fluctuations 
in a relatively sophisticated way; for most theories 
of dynamic heterogeneity, exactly the reverse 
applies. Therefore, neither theory can
claim to offer a complete picture. Much 
evidence on colloids suggests that there are 
indeed localized regions of enhanced 
mobility (e.g. \cite{crocker,weitzcapri}); but 
this very idea requires some sort of 
immobilized background state within 
which such excitations arise. MCT addresses
the onset of this collectively arrested state. 
Ignoring the excitations may be only a modest 
error if the density of excitations 
near the MCT transition is small enough, 
but could undermine the whole approach if 
it is large. Conversely, any theory of 
localized defect dynamics within a 
frozen matrix (modelled, e.g., on a 
lattice \cite{garrahan}) should work only if, 
on coming from the glass side, the 
proliferation of defects is not pre-empted by a 
collective unfreezing of the matrix through an 
MCT-like mechanism.

The adequacy of either theory 
may depend on what type of glass is under study.
For colloids, MCT seems surprisingly adequate 
\cite{goetze2}. A striking recent success concerns 
systems with both attractive interactions and 
hard-core repulsions. First, MCT unambiguously 
predicts 
\cite{bergen,dawson,sciortino} that adding a weak, short range
attraction to the hard sphere system should melt the glass  
(which has $f(q^*)\sim 0.8$). Second, MCT predicts that adding more of
the same attraction should mediate a second arrest, this time into a
`gel' state of high nonergodicity parameter ($f(q^*)\sim
0.95$). Third, MCT predicts that as parameters are varied, a higher
order bifurcation point should enter the picture, resulting in a
characteristic logarithmic decay for $\Phi(q,t)$. Although not every
detail of this scenario is yet confirmed, there is clear experimental
evidence for both the re-entrance, and the logarithmic decay
\cite{pham1,pham3}. The latter is also seen clearly in  
recent simulations \cite{puertas,romesim}. 
These successes of MCT pose a notable challenge to 
dynamical heterogeneity theories of the glass 
transition in colloids. Until such a theory 
can explain the three 
features just outlined, it is fair to conclude 
that MCT remains the least inadequate theory 
of the colloidal glass transition \cite{paris}.

\section{SHEAR THINNING}
\label{thin}

\subsection{A Microscopic Approach}

In Ref.\cite{fuchsprl}, a theory is propounded, along MCT lines, of
colloidal suspensions under flow. The work was intended mainly to
address the case of repulsion-driven glasses, and to study the effect
of imposed shear flow either on a glass, or on a fluid phase very near
the glass transition. In either case, simplifications might be
expected because the bare diffusion time $\tau_0 = a^2/D_0$ is small
compared to the `renormalized' one $\tau = a^2/D$, which in fact
diverges (enslaved to the $\alpha$ relaxation time) as the glass
transition is approached. If the imposed steady shear rate is
$\dot\gamma$, then for $\dot\gamma\tau_0 \ll 1 \leq \dot\gamma\tau$,
one can hope that the details of the local dynamics are inessential
and that universal features related to glass formation should
dominate. Note, however, that by continuing to use a quadratic $H$
(Eq.\ref{hamiltonian}), we will assume that, even under shear, the
system remains `close to equilibrium' in the sense that the density
fluctuations that build up remain small enough for a harmonic
approximation to be useful. This may well be inadequate for hard
spheres, but a systematic means of improvement upon it is not yet
available.

The basic route followed in Ref.\cite{fuchsprl} is quite similar to
that laid out above for standard MCT, modulo the fact that an imposed
shear flow is now present. A key simplification is to neglect velocity
fluctuations so that the imposed shear flow is locally identical to
the macroscopic one; this cannot be completely correct, but allows
progress to be made. For related earlier work see
Refs.\cite{indrani,milner}.

We again take $\beta = 1$, $D_0 = 1$, and start from the Langevin
equations $\dot {\bf r}_i = {\bf u} + {\bf F}_i + {\bf f}_i$ for
independent particles of unit diffusivity subjected to external forces
${\bf F}_i$ and, now, an imposed flow velocity ${\bf u}({\bf
  r}_i)$. We take this to be a simple shear flow with ${\bf u}({\bf
  r}) = \dot\gamma y {\bf \hat x}$. The Smoluchowski equation $\dot
\Psi = \Omega \Psi$ is unchanged but the evolution operator is now
$\Omega = \sum_i\nabla_i.(\nabla_i-{\bf F}_i-{\bf u}({\bf r}_i))$. So
far, the adaption to deal with shearing is fairly trivial. The next
stages are not. We assume an initial equilibrium state with $\Psi(t=0)
\propto \exp[-H]$, and switch on shearing at $t=0+$. We define an {\em
  advected correlator} 
\beq
\Phi({\bf q},t) = \langle 
\rho({\bf q},0)\rho(-{\bf q}(t),t)\rangle/S(q)N
\label{advec}
\eeq
where ${\bf q}(t) = (q_{x} , q_{y} + q_{x} \gd t, q_{z})$.
%{\bf q} + {\bf q}.{\bf K} t$ with ${\bf K}$ the
%velocity gradient tensor, $K_{ij} = \dot\gamma\delta_{ix}\delta_{jy}$. 
This definition subtracts out the
trivial part of the advection, which merely transports density
fluctuations from place to place. The nontrivial part comes from the
effect of this transport on their time evolution; the main effect (see
e.g. \cite{milner}) is to kill off fluctuations by moving their
wavenumbers away from $q^*$ where restoring forces are weakest (hence
the peak there in $S(q)$). Hence the fluctuations feel a stronger
restoring force coming from $H$, and decay away more strongly. This
feeds back, through the nonlinear term, onto the other fluctuations,
including ones transverse to the flow and its gradient (i.e., with
${\bf q}$ along $z$) for which the trivial advection is absent.

There follow a series of MCT-like manipulations which differ from
those of the standard approach because they explicitly deal with the
switching on of the flow at $t=0+$. We integrate through the transient
response to obtain the steady state correlators, under shear, as
$t\to\infty$. (There is no integration through transients in standard
MCT; one works directly with steady-state quantities.) Despite all
this, the structure of the resulting equations is remarkably similar
to Eqs. \ref{correlator},\ref{memory}: 
\beq
\dot\Phi({\bf q},t) + \Gamma({\bf q},t)\left[\Phi({\bf
    q},t)+\int_0^tm({\bf q},t,t')\dot\Phi({\bf q},t')dt'\right] = 0 
\label{corr1}
\eeq

This equation describes transient 
density fluctuations; that is, time zero 
in Eq.\ref{advec} corresponds to the switch-on time
of a shear flow, so that $\Phi$ is a particular
instance of a two-time correlator rather than
a one-time one. 
%{\bf Matthias: please confirm}. MF: Is correct.
For a related approach based on the one-time
correlator of fluctuations
around steady state (but assuming a fluctuation-dissipation
theorem which need not apply here), see \cite{Miya}.
Eq.\ref{corr1} involves a time dependent, 
anisotropic ``initial decay rate'':
\beq
\Gamma({\bf q},t)S(q) = q^2 + q_xq_y\dot\gamma t+ (q_xq_y\dot\gamma t+
q_x^2\dot\gamma^2 t^2)S(q) - q_xq_y\dot\gamma S'(q)/q 
\eeq
The memory kernel is no longer a function of the time interval $t-t'$
but depends on both arguments separately 
\beq
m({\bf q},t,t')= \sum_{{\bf k}} V({\bf q},{\bf k},t,t')\Phi({\bf
  k},t)\Phi({\bf k-q},t) 
\eeq
through a time-dependent vertex $V$ whose detailed derivation will
appear in \cite{fuchsvertex}:\label{vertsec} 
\begin{eqnarray}
& V({\bf q},{\bf k},t,t')
= \left[ N S_k S_p / \left( 2 V^{2} S_q \Gamma({\bf q},t) \Gamma({\bf q},t') 
\right)  \right] & \nonumber \\ &
\left[ \, {\bf q}(t).{\bf k}(t\!\!-\!\!t') c(k(t\!-\!t'))
+ {\bf q}(t).{\bf p}(t\!\!-\!\!t') c(p(t\!-\!t')) + \right. 
& \nonumber \\ & \left.
S(q(t\!-\!t')) {\bf q}(t).\left( {\bf q}(t\!\!-\!\!t')
c(q(t\!-\!t')) \!\!-\!\! {\bf q} c(q) 
\right) \, \right]  & \nonumber \\ & 
\left[ \, {\bf q}.{\bf k} c(k)
+ {\bf q}.{\bf p} c(p) + \right. 
& \nonumber \\ & \left.
q_{x} q_{y} \gd t \, S(q) \left(
c(k) + c(p) - \frac{N}{2V} ( c(q) c(k) + c(q) c(p) + c(k) c(p) ) 
\right) \, \right] 
 &\label{vertexformula}
\end{eqnarray}
Here  the abbreviation $\bf p = \bf q - \bf k$ is used, and 
advected wavevectors (see below Eq. \ref{advec}) carry a time dependence.
%***GIVE HERE THE LENGTHY VERTEX THAT EVERYONE IS WAITING FOR, BUT
%   WITHOUT DERIVATION***** 

Using a nonequilibrium Kubo-type relationship \cite{fuchsprl} one can
also obtain an expression for the steady state viscosity $\eta =
\sigma(\dot\gamma)/\dot\gamma$ where $\sigma(\dot\gamma)$ is the shear
stress as a function of shear rate. The viscosity is expressed as an
integral of the form 
\beq
\eta = \int_0^\infty dt \sum_{{\bf k}} V_\eta({\bf k},t) \Phi^2({\bf k},t)
\label{kubolike}
\eeq
where the function $V_\eta$ may be found in Ref. \cite{fuchsprl}.

The above calculations give several interesting results. 
First, any nonzero shear rate, however small, restores ergodicity for
all wavevectors (including ones which are transverse to the flow and
do not undergo direct advection). This is important, since it is the
absence of ergodicity that normally prevents MCT-like theories being
used inside the glass phase, at $T<T_g$ or $\phi>\phi_g$. Here we may
use the theory in that region, so long as the shear rate is finite.

In the liquid phase ($\phi<\phi_g$) the resulting flow curve
$\sigma(\dot\gamma)$ shows shear thinning at $\dot\gamma\tau
\gtrapprox 1$, which is when the shearing becomes significant on the
timescale of the slow relaxations. This is basically as expected. Less
obviously, throughout the glass, one finds that the limit
$\sigma(\dot\gamma \to 0+) \equiv \sigma_Y$ is nonzero. This quantity
is called the yield stress and represents the minimum stress that
needs to be applied before the system will respond with a steady-state
flow. (For lower stresses, various forms of creep are possible, but
the flow rate vanishes in steady state.)

The prediction of a yield stress in colloidal glasses is significant,
because glasses, operationally speaking, are normally defined by the
divergence of the viscosity. However, it is quite possible for the
viscosity to diverge without there being a yield stress, for example
in `power law fluids' where $\sigma(\dot\gamma) \sim \dot\gamma^p$
with $0<p<1$ \cite{sgr}.  
This does not happen in the present calculation, where the yield
stress jumps discontinuously from zero to a nonzero value,
$\sigma_Y^c$, at $\phi_g$. The existence of a yield stress seems to be
in line with most experimental data on the flow of colloidal glasses,
although one must warn that experimentalists definitions of what a
yield stress is, do vary across the literature \cite{barnes}. Ours is
defined as the limiting stress achieved in a sequence of experiments
at ever decreasing $\dot\gamma$, ensuring that {\em a steady state is
  reached} for each shear rate before moving onto the next one. The
latter requirement may not be practically achievable since the
equilibration time could diverge smoothly at small $\dot\gamma$:
certainly one would expect to have to wait at least for times $t$ such
that $\dot\gamma t \gtrapprox 1$. But unless the flow curve has
unexpected structure  
(absent in this approach) at small shear rates, the required
extrapolation can presumably be made.

\subsection{Schematic MCT models}

\begin{figure}
\centerline{\psfig{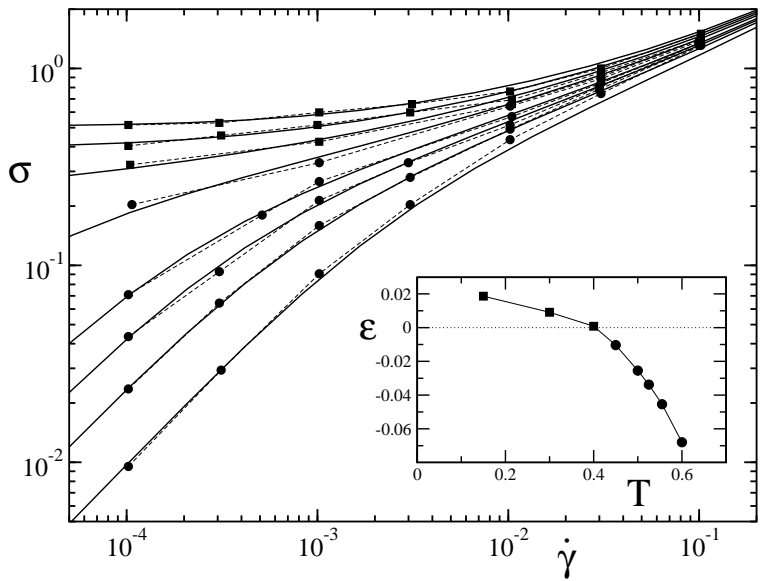}}
\caption{\label{newfig}Symbols are shear stress 
 data of a super--cooled Lennard--Jones binary mixture in
reduced units taken from Ref. \cite{Berthier02}; from top to bottom,
the temperatures are  0.15, 0.3, 0.4, 0.45, 0.5, 0.525, 0.555 \& 0.6
while the transition lies at
$T \approx 0.435$. The solid lines give fits by eye using the
F$_{12}^{\gd}$--model with parameter $\varepsilon$ measuring the separation
from the transition shown in the inset.
Units are converted by $\sigma=1.5\sigma^{\rm theo.}$ and 
$\gd=1.3\gd^{\rm theo.}$. More details and a preliminary 
fit can be found in \cite{JPCM}.}
\end{figure}

It has long been known that the key mathematical structure behind
Eqs. \ref{correlator}-\ref{vertex} can be captured by low-dimensional
schematic models in which the full ${\bf q}$ dependence  
is suppressed \cite{goetze85,goetze}. In other words, one chooses a
single mode, with a representative wavevector around the peak of the
static structure factor, and writes mode coupling equations for this
mode treated by itself. At a phenomenological level, one can capture
the physics similarly even with shearing present (despite the more
complicated vectorial structure that in reality this
implies). Specifically one can define \cite{fuchsprl} the
$F_{12}^{\dot\gamma}$ model --- the sheared extension of a well known
static model, $F_{12}$ --- via 
\beq
\dot\Phi(t) + \Gamma \left[ \Phi(t) + \int_0^t m(t-t')\dot\Phi(t')
  dt'\right] = 0 
\label{scheme1}\eeq
with memory function (schematically incorporating shear)
\beq
m(t) = [v_1\Phi(t)+v_2\Phi^2(t)]/(1+\dot\gamma^2t^2)
\label{scheme2}\eeq
The vertex parameters $v_{1,2}$ are smooth functions of the volume
fraction $\phi$ (and any interactions). 
To calculate flow curves, etc., one also needs a schematic form of
Eq.\ref{kubolike}; here we take the first moment of the correlator to
fix the time scale for stress relaxation (which is, in suitable units,
simply the viscosity): 
\beq
\eta = \int_0^\infty \Phi(t) dt
\label{scheme3}\eeq
(Note that a different choice, e.g. with $\Phi(t)^2$ in this equation
to closer resemble Eq.\ref{kubolike}, would yield quite similar
results.) 
This simplest of schematic models gives very 
similar results to a much more sophisticated (but still
schematic) approximation of the full equations \cite{fuchsprl}, with
$\sigma-\sigma_Y\sim \dot\gamma^{0.16}$ and
$\sigma_Y-\sigma_Y^c\sim(\phi-\phi_g)^{1/2}$. 
Such predictions can be compared with experiment
\cite{JPCM} and, as shown in Figure \ref{newfig},
suggest that the more advanced schematic models
are at least semi-quantitative.

%{\bf add {newfig} here, plus caption -- from JPCM or
%else more refined fit. Caption must refer to JPCM
%for details of the advanced schematic models (there's
%not room for details here)}

\section{SHEAR THICKENING AND JAMMING}

\begin{figure}[h]
\centerline{\psfig{file=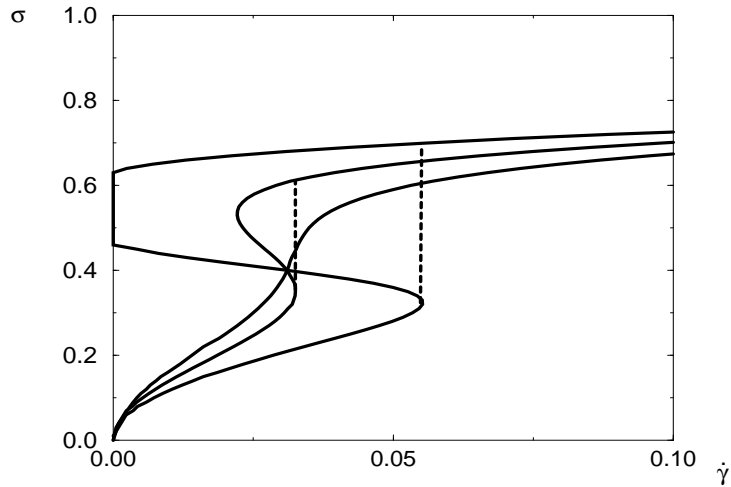,width=10.cm}}  
\caption{Three possible flow curves for a shear thickening
  material. The monotonic curve corresponds to continuous shear
  thickening. The remaining two curves are S-shaped; one expects, on
  increasing the shear rate, the stress to jump from the lower to
  upper branch at (or before) the vertical dashed line shown in each
  case. One curve shows the full jamming scenario: the existence of an
  interval of stress, here between 0.45 and 0.63, within which the
  flow rate is zero, even in a system ergodic at rest. (Stress and
  strain rate units are arbitrary.)}  
\label{fig4}
\end{figure}

The calculations described above predict, generically, shear thinning
behaviour: advection kills fluctuations, reducing the $\alpha$
relaxation time, which causes the system to flow more easily at higher
stresses. However, in some colloidal systems, the reverse occurs. This
is shear thickening, and gives a flow curve $\sigma(\dot\gamma)$ with
upward curvature. In extreme cases, an essentially vertical portion of
the curve is reported \cite{Laun,Bender}. One interpretation of the
latter scenario (called `discontinuous shear thickening') is that the
underlying flow curve is actually S-shaped. Since any part of the
curve with negative slope is mechanically unstable (a small increase
in the local shear rate would cause an acceleration with positive
feedback), this allows a hysteresis cycle in which, at least according
to the simplest models, discontinuous vertical jumps on the curve
bypass the unstable section (see Figure \ref{fig4}).

If this viewpoint is adopted, there seems to be nothing to prevent the
upper, re-entrant part of the curve from extending right back to the
vertical axis (see Figure \ref{fig4}) in which case there is zero
steady-state flow within a certain interval of stress. The system has
both an upper and a lower yield stress delimiting this region. (If it
is nonergodic at rest, it could also have a regular yield stress on
the lower part of the curve near the origin -- we ignore this here.)
This case has been called `full jamming' \cite{head}. Although mostly
a theoretical speculation, at least one experimental report of this
kind of behaviour has appeared in the literature recently
\cite{bibette}.

The above discussion suggests that shear thickening and full jamming
might be viewed as a stress-induced glass transition of some sort
\cite{jamming}. If so, it is natural to ask whether this idea can be
accommodated within an MCT-like approach. 
Since the analysis of Ref. \cite{fuchsprl} gives only shear thinning,
this is far from obvious. In particular, a stress-induced glass
transition would require the vertex $V$ to `see' the stress; this
might require one to go beyond harmonic order in the density, that is,
it might require improvement to Eq.\ref{hamiltonian}. Indeed, since it
is thought that jamming arises by the growth of chainlike arrangements
of strong local compressive contacts \cite{jamming}, it is very
reasonable to assume that correlators beyond second order in density
should enter.

In \cite{epl} we develop a schematic model along the lines of
Eqs.\ref{scheme1}--\ref{scheme3} to address shear thickening (with,
for simplicity, $v_2=0$). This is the $F_{1}^{\dot\gamma,\sigma}$
model
\beq 
\dot\Phi(t) + \Gamma \left[ \Phi(t) + \int_0^t m(t-t')\dot\Phi(t')
  dt'\right] = 0 
\label{colinscheme1}\eeq
with memory function
\beq
m(t) = [v_0+\alpha\sigma]\exp[-\dot\gamma t]\Phi(t) 
\label{colinscheme2}\eeq
and viscosity $\eta = \sigma/\dot\gamma$ obeying 
\beq
\eta = \int_0^\infty \Phi(t) dt\,.
\label{colinscheme3}\eeq
The memory function now schematically incorporates both the loss of
memory by shearing and a stress-induced shift of the glass
transition. (Without stress or shear, the latter occurs at $v_0 = 4$.)
The choice of an exponential strain rate dependence is purely for
algebraic convenience, whereas the form in Eq. \ref{scheme2} is closer
to the one found in the full ${\bf q}$-dependent vertex under shear
(see section \ref{vertsec} above and \cite{fuchsprl}). The choice of a
linear dependence of the vertex on stress (rather than the quadratic
one that would arise in a Taylor expansion about the quiescent state)
can be viewed as a linearization about a finite stress chosen to lie
close to the full jamming region: this, rather than the behaviour at
very small stresses, is the interesting region of the model. In any
case, the qualitative scenarios that emerge from
Eqs.\ref{colinscheme1}--\ref{colinscheme3} are relatively robust to
the precise details of the model \cite{epl}.

\begin{figure}
\centerline{\psfig{file= 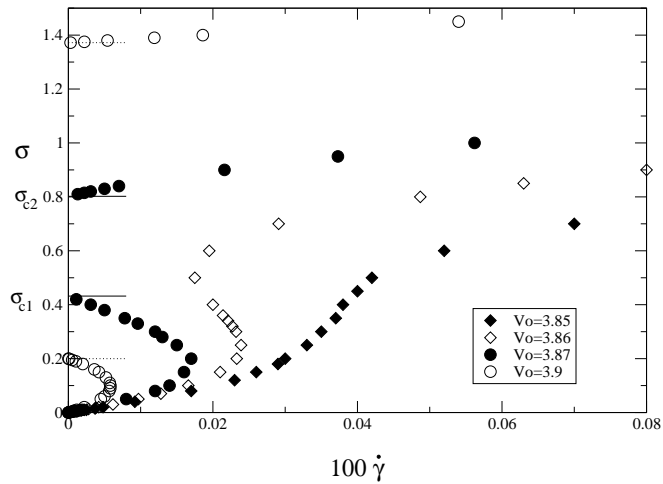,width=10.cm}}  
%\onefigure[scale=0.4]{FlowcurvesAlpha=0.95.eps}
\caption{Flow curves for $\alpha=0.95$. For the two largest values of $v_o$,
it appears that for a window in $\sigma$, the relaxation time has
diverged. Analytic calculations of the limits of this window are
indicated as horizontal lines near the stress axis. These values of
the stress are dubbed $\sigma_{c1}$ and $\sigma_{c2}$, as
shown here for one of the parameter sets. 
%For Vo=3.9: jam between sigma = 0.200 and 1.372;
%For Vo=3.87:  ``    ``   ``     0.432     0.802
}\label{flowcurves}
\end{figure}

This model results in a `full jamming'  scenario as part of a wider
range of rheological behaviour. Fig. \ref{flowcurves} shows three
kinds of thickening behaviour, dependent on model parameters; $v_0$ is
varied close to the quiescent glass transition, and for the chosen
$\alpha$ there is a progression from a monotonic, continuously
shear-thickening curve, via a nonmonotonic S-shaped curve, to a curve
that extends right back to the vertical axis. For the largest values
of the parameter $v_o$, in Fig. \ref{flowcurves}, 
there is therefore a range of stress for which the shear rate returns to zero:
there is then no ergodic solution, and the jammed state is
stable. This represents full jamming. Note that if, as seems likely,
$\alpha$ depends on the details of interparticle interactions, then
the evolution between these scenarios does too. This makes sense since
one would certainly expect hard particles to be more `jammable' than
soft ones.

Fig.\ref{flowcurves} is qualitatively similar to Fig.\ref{fig4} --
whose data actually comes from \cite{head}. In that work a somewhat
similar theory is developed, based not on MCT but on the trap model of
glasses. The emergence of the same qualitative scenario from two quite
different approaches to glass rheology is reassuring, although in each
case the ansatz of a stress-dependent glass transition was,
effectively, put in by hand.

\begin{figure}

\centerline{\psfig{file= 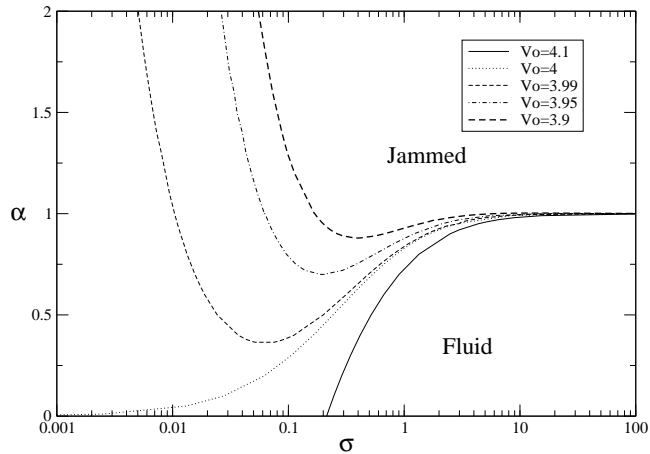,width=10.cm}}  

%\onefigure[scale=0.4]{pdiagram.eps}

\caption{`Phase diagrams' for the model for various $v_o$. The lines
  denote transitions in the $(\alpha,\sigma)$-plane. All states below
  the curve for a given value of $v_o$ are fluid states, whilst those
  above (and on) the line are nonergodic, jammed
  states.}\label{pdiagram}  

\end{figure}

The lower and upper endpoints $\sigma_{c1}$ and $\sigma_{c2}$ of the
stable jammed state represent distinct jamming transitions. Their
critical stresses obey
\begin{equation}
f_c\left[\left(v_o+\alpha\sigma_c\right)f_c-2\right]=\sigma_c,
\label{transition}  
\end{equation}
where $f_c$ is given by the largest solution of
$\frac{f_c}{1-f_c}=(v_o+\alpha\sigma_c)f_c^2$ . 
Such transitions exist provided that both $v_o$ and $\alpha$ are
sufficiently large. Bertrand {\it
et al} \cite{bibette} found that, for concentrations
below a certain value, their samples showed ordinary thickening,
whilst above this 
value the shear-induced solid was seen. The behaviour illustrated in
Fig. \ref{flowcurves} is reminiscent of this. Note that the
re-fluidisation under increasing stress depends on 
$\alpha$: if this is too large (for a given
$v_o$) this re-fluidisation is not present. The resulting `phase
diagram' of the model, dividing parameter space into ergodic 
and nonergodic regions, is shown in Fig. \ref{pdiagram}. At large
enough stresses, jammed states arise for $\alpha > 1$. However, for
particle densities close to but below the quiescent glass transition,
for $\alpha<1$ the system jams in an intermediate window of stress.

\section{GLASSY VERSUS HYDRODYNAMIC THICKENING}
\label{hydro}

Shear thickening is widely reported
(e.g. \cite{Laun,Bender,Frith,O'Brien+Mackay}) and usually attributed
to a buildup of hydrodynamic forces between clusters of particles
\cite{BradyReview,melrosecluster}. Our work suggests that, at least in
some systems, this may not be the only mechanism at work. In
particular, Fig. \ref{flowcurves} admits shear thickening at Peclet
numbers $\dot\gamma\tau_0 \sim 10^{-4}$, rather than values of order
unity predicted by most theories of hydrodynamic clustering. Such
theories do not so far appear to offer any natural explanation of the
S shaped flow curve that appears to underly {\em discontinuous} shear
thickening (see e.g. \cite{bergbrady}, and references therein). On the
other hand, simulations of dense colloids do predict, for hard spheres
in the absence of Brownian motion, a catastrophic jamming
transition. In this transition, a network of close contacts propagates
to infinity at {\em finite strain}, creating a solid
\cite{melroseball}. The relation between this and our own model in
which Brownian motion of course plays an essential role, is yet to
become clear.

It should also be pointed out that, to whatever extent full jamming is
actually observed \cite{bibette}, hydrodynamic theories cannot explain
it. This is because hydrodynamic forces are dynamical in origin and
therefore cannot be responsible for maintaining a purely static state
of arrest. It is however conceivable that a limit exists in which
interparticle velocities and separations both vanish at late times in
such a fashion that the resulting forces approach constant
values. However, we do not find this particularly plausible.

Note in any case that existing hydrodynamic theories (rather than
simulations \cite{melroseball}) of colloid rheology, by taking no
account of the glass transition, predict that the zero shear viscosity
diverges only at random close packing (volume fraction $\phi = 0.63$)
\cite{bradytheory}. This appears inconsistent with experimental
observations where the viscosity divergence occurs instead at the
colloidal glass transition ($\phi = \phi_g = 0.58$)
\cite{pooneta}. Accordingly it is necessary to develop a new theory,
as outlined in Section \ref{thin}, to describe flow curves at $\phi >
\phi_g$. The hypothesis of the work on colloidal jamming reported here
is that the proximity of this transition also affects flow properties
in a window of densities {\em below} $\phi_g$, to the extent that one
should treat hydrodynamic forces as a perturbation to the dynamics of
collective arrest, rather than vice versa.

\section{CONCLUSION}

Mode Coupling Theory (MCT) has had important recent successes, such as
predicting, in advance of experiment, the re-entrant glass/gel
nonergodicity curves that arise in colloidal systems with short range
attractions \cite{bergen,dawson,sciortino,pham1}.

Theoretical developments directly inspired by MCT now offer a
promising framework for calculating the nonlinear flow behaviour of
colloidal glasses and glassy liquids \cite{fuchsprl}. In fact, this
offers the only current prospect for {\em quantitative} prediction of
yield behaviour and nonlinear rheology in this or any other class of
nonergodic soft materials. (Other work on the rheology of glasses
\cite{sgr,bbk} does not, as yet, offer quantitative prediction of
experimental quantities.) While promising, many things are missing so
far from the approach initiated in \cite{fuchsprl}: velocity
fluctuations, hydrodynamic forces, anharmonicity in $H$ etc., are all
ignored. The fact that only shear thinning is predicted in this case
is excusable.

The schematic work of Ref.\cite{epl} on shear thickening is
preliminary, but interesting in that it suggests how new physics
(beyond two-point correlations) may need to be added to MCT before the
full range of observed colloidal flow behaviour is properly
described. Hydrodynamic interactions, and perhaps velocity
fluctuations, are certainly also important in some aspects of shear
thickening, as discussed in Section \ref{hydro}, though we might hope
that these do not dominate very close to the glass transition where
the longest relaxation time is structural rather than hydrodynamic. Of
course, even for systems at rest, it is known that some important
physics is missing from MCT, in particular, the kinds of activated
dynamics discussed in Section \ref{assisted}. These allow the system
to move exponentially slowly despite being in a region of phase space
where, according to MCT, it cannot move at all (see
e.g. \cite{kk}). Qualitatively, stress-induced jamming seems a quite
different phenomenon from this, although one cannot rule out a link of
some sort (e.g. if stress switches off the activated processes
\cite{head}). Accordingly we can suspect that there are more things
missing from MCT than just activated processes. In particular a more
general treatment of anharmonic terms (or equivalently, a treatment of
three-point and higher order correlations) may be required before one
has a fully workable theory of sheared colloidal glasses.


\begin{thebibliography}{99}
\bibitem{book} Cates M. E. and Evans M. R., ``Soft and Fragile Matter:
  Nonequilibrium Dynamics, Metastability and Flow'' IOP Publishing,
  Bristol (2000). 

\bibitem{pine} Pine D. J., ``Light Scattering and Rheology of Complex
  Fluids Driven far from Equilibrium'', in \cite{book}, pp.9--47. 

\bibitem{kob} Kob W., ``Supercooled Liquids and Glasses'', in
  \cite{book}, pp. 259--284. 

\bibitem{poon} Poon W. C. K., Starrs L., Meeker S. P., Moussaid A.,
  Evans R. M. L., Pusey P. N. and Robins M. M., Faraday Discuss. 112
  (1999) 143--154. 

Poon W. C. K., Renth F., Evans R. M. L., Fairhurst D. J., Cates
M. E. and Pusey P. N., Phys .Rev. Lett. 83 (1999) 1239--1243. 

\bibitem{kroy} Kroy K., et al., work in progress.

\bibitem{goetze}G\"otze W. and Sjoegren L., Rep. Prog. Phys. 55
  (1992), 241--376. 

\bibitem{ramaswamy} Ramaswamy S., ``Self-Diffusion of Colloids at
  Freezing'', in ``Theoretical Challenges in the Dynamics of Complex
  Fluids,'', McLeish T. C. B., Ed., pp7-20, Kluwer, Dordrecht 1987.   

\bibitem{kk} Kawasaki K. and Kim B., J. Phys. Cond. Mat. 14 (2002) 2265--2273.

\bibitem{paris} Cates M. E., cond-mat/0211066, to appear in Ann. Henri
  Poincare. 

\bibitem{garrahan} Garrahan J. P. and Chandler D., Phys. Rev. Lett. 89
  (2002) 035704; J. Garrahan, this volume. 

\bibitem{crocker} Weeks E. R., Crocker J. C., Levitt A. C., Schofield
  A., Weitz D. A., Science 287 (2000) 627. 

\bibitem{weitzcapri} Weitz D. A., this volume.

\bibitem{goetze2} G\"otze W., ``Aspects of Structural Glass
  Transitions'', in ``Liquids, Freezing and Glass Transition'', Hansen
  J. P., Levesque D., and Zinn-Justin J. Eds (Les Houches Session LI),
  North Holland 1991, pp. 287--503. van Megen W., Underwood S. M. and
  Pusey P. N., Phys. Rev. Lett. 67 (1991) 1586--1589; van Megen W. and
  Underwood S. M., Phys. Rev. E. 49 (1994) 4206--4220;
  G\"otze W., J. Phys.: Cond. Matt. 11, A1-A45 (1999). 

\bibitem{bergen} Bergenholtz J., and Fuchs, M., Phys. Rev. E 59 (1999)
  5706--5715. 

\bibitem{dawson} Dawson K., Foffi G., Fuchs M., Gotze W., Sciortino
  F., Sperl M., Tartaglia P., Voigtmann T., Zaccarelli E.,
  Phys. Rev. E 63 (2001) 011401. 

\bibitem{sciortino} Fabbian L., G\"otze W., Sciortino F., Tartaglia
  P., and Thiery F., Phys. Rev. E 59 (1999) R1347--R1350. 1999;
  Sciortino F., this volume, Sciortino F., Nature Materials 1 (2003)
  145--146.  

\bibitem{pham1} Pham K. N., Puertas A. M., Bergenholtz J., Egelhaaf
  S. U., Moussaid A., Pusey P. N., Schofield A. B., Cates M. E., Fuchs
  M. and Poon W. C. K., Science 296 (2002) 104--106. Poon W. C. K.,
  Pham K. N., Egelhaaf S. U. and Pusey P. N., J. Phys. Cond. Matt. 15
  (2003) S269-S275. 

\bibitem{pham3} Pham K. N., private communication.

\bibitem{puertas} Puertas A. M., Fuchs M. and Cates M. E.,
  Phys. Rev. Lett. 88 (2002) 098301. 

\bibitem{romesim} Sciortino F., 
Tartaglia P., Zaccarelli E., 
cond-mat/0304192.

\bibitem{fuchsprl} Fuchs M. and Cates M. E., Phys. Rev. Lett. 89
  (2002) 248303;  Faraday Discussion 123 (2002) 267--286.  

\bibitem{indrani} Indrani A. V. and Ramaswamy S., Phys. Rev. E  52
  (1995)  6492--6496. 

\bibitem{milner} Cates M. E. and Milner S. T., Phys. Rev. Lett. 62
  (1989)  1856--1859. 

\bibitem{Miya} Miyazaki K., Reichman D. R., Phys. Rev.
E 66 (2002) 050501(R).

\bibitem{fuchsvertex} Fuchs M. and Cates M. E., in preparation.

\bibitem{bouchaud} Bouchaud J.-P., J. Physique I 2 (1992) 1705--1713.

\bibitem{sgr} Fielding S. M., Sollich P. and Cates M. E., J. Rheol. 44
  (2000) 323--369; Sollich P., Lequeux, F., Hebraud P. and Cates
  M. E., Phys. Rev. Lett. 78 (1987) 2020--2023. Sollich P.,
  Phys. Rev. E 58 (1998) 738--759. 

\bibitem{barnes} Barnes H. A. Hutton J. F. and Walters, K., ``An
  introduction to rheology'', Elsevier, Amsterdam 1989. 

\bibitem{goetze85} G\"otze W., Z. Phys. B 60 (1985) 195.

\bibitem{JPCM} Fuchs M., Cates M. E., 
J. Phys. Cond. Mat. 15 (2003) S401.

\bibitem{Berthier02}
Berthier L. and Barrat J.~L., J. Chem. Phys. 116 (2002) 6228.
\bibitem{Laun} Laun H. M., J. Non-Newtonian Fluid Mec. 54 (1994) 87--108.


\bibitem{Bender} Bender J. and Wagner N. J., J. Rheol. 40 (1996) 889--916. 

\bibitem{head} Head D. A., Ajdari A. and Cates M. E., Phys. Rev. E 64
  (2001) 061509. 

\bibitem{bibette} Bertrand E., Bibette J. and Schmitt V., Phys. Rev. E
  66 (2002) 06040(R). 

\bibitem{jamming} Cates M. E., Wittmer J. P., Bouchaud J.-P. and Claudin P., 
Phys. Rev. Lett. 81 (1998) 1841--1844. Liu A. J. and Nagel S. R.,
Nature 396 (1998) 21--22. Ball R. C. and Melrose J. R., Adv. Colloid
Interface Sci. 59 (1995) 19--30. 

\bibitem{epl} Holmes C., Fuchs M. and Cates M. E., Europhys. Lett. 63
  (2003) 240--246. 

\bibitem{Frith}
Frith W. J., d'Haene P., Buscall R. and Mewis J.,
J. Rheol{40} (1996){531}.

\bibitem{O'Brien+Mackay} O'Brien V. T. and Mackay M. E.,
Langmuir {16}(2000){7931}.

\bibitem{BradyReview}
Brady J. F.,
Curr. Opin. Colloid Interface Sci. {1} (1996) {472}, and references therein.

\bibitem{melrosecluster} Farr, Melrose, Ball Phys. Rev. E 55 (1997) 7203--7211.

\bibitem{bergbrady} Bergenholtz J., Brady J. F., Vicic M., J. Fluid
  Mech. 456 (2002) 239--275. 

\bibitem{melroseball}
Ball R. C. and Melrose J. R.,
Adv. Colloid Interface Sci. {59}(1995){19}.

\bibitem{bradytheory} Brady J. F., J. Chem. Phys. 91 (1993), 3335--3341.

\bibitem{pooneta} Segre P. N., Meeker S. P., Pusey P. N. and Poon
  W. C. K., Phys. Rev. Lett. 75 (1995) 958--961;
Cheng Z., Zhu J., Chaikin P. M., Phan S.-E., and Russel W. B.,
 Phys. Rev. E 65, 041405 (2002). 

\bibitem{bbk} Berthier L., Barrat J.-L. and Kurchan J. Phys. Rev. E 61
  (2000) 5464--5472.  

\end{thebibliography}
\end{document}